# Computational Essays: An Avenue for Scientific Creativity in Physics

Tor Ole B. Odden

*Center for Computing in Science Education, Department of Physics, University of Oslo, 0316 Oslo, Norway*

Marcos D. Caballero

*Center for Computing in Science Education, Department of Physics, University of Oslo, 0316 Oslo, Norway and
Department of Physics and Astronomy and Create for STEM Institute, Michigan State University, East Lansing, MI 48823*

Computation holds great potential for introducing new opportunities for creativity and exploration into the physics curriculum. At the University of Oslo we have begun development of a new class of assignment called computational essays to help facilitate creative, open-ended computational physics projects. Computational essays are a type of essay or narrative that combine text and code to express an idea or make an argument, usually written in computational notebooks. During a pilot implementation of computational essays in an introductory electricity and magnetism course, students reported that computational essays facilitated creative investigation at a variety of levels within their physics course. They also reported finding this creativity as being both challenging and motivating. Based on these reflections, we argue that computational essays are a useful tool for leveraging the creative affordances of programming in physics education.

## I. INTRODUCTION

Traditionally, learning physics is not an especially creative endeavor. In many physics courses, students spend most of their time solving problems that have already been solved hundreds if not thousands of times, which have (at most) one correct answer, and which have only a small number of viable paths to the solution. Laboratory work is often distilled into a set of prescribed steps, with little room for deviation or exploration. This educational model, however, is not authentic to the discipline of physics—outside of classroom contexts there are seldom singular correct answers, and there is almost never just one path to a solution. Often, "real" physics is a process of refinement, as expressed by the maxim "nothing works the first time" [1].

Certain reform-based teaching practices have started to add opportunities for creativity back into the curriculum. For example, design-based labs present students with challenging, open-ended tasks that have multiple possible solutions [2–4]. Modeling curricula have also opened up room for alternative approaches to problem-solving [5–7], aided in large part by the increased integration of computation into physics education. However, more remains to be done in this area.

Computation offers numerous affordances for this type of creative learning. Computation allows students to accomplish much with a relatively small set of programming techniques. This can be empowering to some students, and opens up opportunities for exploration-based learning [8]. Additionally, computation allows students to easily explore and "play around" with topics or concepts that might be analytically intractable or far above their educational level [9]. However, these creative affordances of programming have not been well developed or studied within the context of physics education. In this study we examine how computation can create opportunities for creativity in the physics curriculum, and how students experience that creative freedom. Specifically, we use a newly-developed class of assignment called a computational essay to explore the ways in which programming can open up creative opportunities for physics learning.

## II. COMPUTATIONAL ESSAY DESIGN PRINCIPLES

Terms like *creativity* can be difficult to operationalize since they tend to have so many meanings across different contexts. In this study we are specifically interested in *scientific creativity* in the context of task and curriculum design (rather than as a property of individual students).

### A. A framework for scientific creativity in education

Drawing on the limited work explicitly addressing creativity in the science education and physics education literature [10, 11], we gauge the scientifically creative affordances of tasks and curricula by looking for the following qualities:
1. Openness in task design
2. Opportunities for original solutions
3. Opportunities for exploration
4. Opportunities for collaboration and cooperation
5. Disciplinary authenticity

Based on this framework, we expect that tasks which present opportunities for creativity are those which give students a challenge that is open-ended, in that there is no single right answer nor an obvious singular path forward. Such challenges should be related and relevant to the physics discipline, while also being accessible enough that students are able to "play around" and explore a variety of fruitful ideas related to possible solutions. They should also offer students the possibility to collaborate with one another, both when working on the solution and when presenting their results.

As previously stated, we see significant potential to leverage computation and programming to create these types of tasks. Therefore, our research questions are as follows: *How can scientific creativity be used as a design principle to create open-ended computational projects? What effect does this project design have on student experiences?*

### B. Computational essay project context and design

To address these questions, we are developing a new type of computational assignment in a computation-rich physics learning environment, the physics department of the University of Oslo, Norway. Since 2003 computation has been a cornerstone of the University of Oslo physics major, with all students taking a Python programming course and a numerical methods course during their first semester. Subsequent courses build on this programming foundation, having students write simulations as part of their weekly homework assignments and exams. Building on this existing course structure, in 2018 we began development of a teaching tool intended to capitalize on this programming focus. We called this tool a computational essay.

Computational essays were originally proposed by diSessa [9] as a form of writing that uses text, along with small programs, interactive diagrams, and computational tools to express an idea. They are often written in so-called notebook environments such as Jupyter notebooks [12], programming environments that allow users to combine code and text into a single document. Notebooks consist of a series of blanks (called "inputs") in which users can type both code and text, from single lines to whole programs or paragraphs. They have long been used by data scientists and professional physicists, both in exploratory analysis and presenting findings [13, 14]. However, to our knowledge Jupyter notebooks and computational essays have not yet been widely used in educational environments.

In the current project, we conceptualize computational essays as a type of essay or report that explicitly incorporates live code to support its thesis, usually written in a notebook environment. Computational essays include all of the elements one would expect in an ordinary essay: for example, an introduction, thesis statement, body paragraphs, and conclusion. They also have a similar purpose, to present a step-by-step argument or explanation. However, the argument in a computational essay is driven by the output of various blocks of code, with the text serving to both explicate the meaning of the code and to explain the output (see Fig. 1).

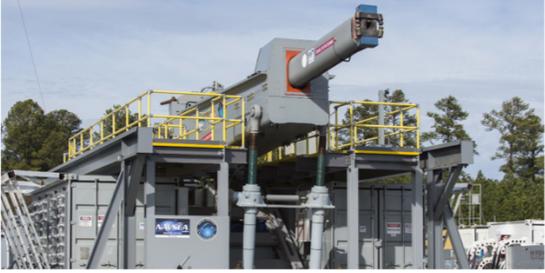

FIG. 1. Example computational essay, showing the mix of text, code, and pictures.

In the fall semester of 2018, we ran a small pilot project to explore how computational essays could be used in the University of Oslo's introductory undergraduate E&M course for physics, engineering, and geoscience majors (roughly 270 enrolled students). Students in the course were given the option to write a computational essay as an alternative to a mandatory presentation-based assignment. Those who chose to participate were challenged to conduct an open-ended computational investigation of some phenomenon related to the course content, then to write a computational essay that summarized their investigation. Students were allowed to work individually or in pairs, and were given approximately 4-6 weeks to conduct their investigations. At the end of the semester students presented their essays to their peers in mock research-group meetings, and the resulting essays were graded pass/fail.

To help scaffold the assignment, we provided students with a project description outlining the expectations, and an example computational essay on the topic "how much current would one need to use a railgun to resupply the International Space Station?" (Fig. 1). We also provided the students with several "basic" simulations of electricity and magnetism phenomena written in Jupyter notebooks, including a simulations of a cyclotron, storm cloud, lightning strike, and magnetic trap [15] for the students to build on if they chose, as well as some suggested investigation questions. The first author also staffed "help desk" hours where students could drop in to ask questions or work on projects.

To address our first research question, this project was designed to explicitly incorporate the above-mentioned aspects of scientific creativity. The project had an open-ended task design, in that students were encouraged to define and pursue investigation questions based on their own interests. Because students defined their own questions, their solutions (i.e., the investigative approaches, modified or re-written code, and explanatory text) were also original. This open-endedness was also tied to significant opportunities for exploration, in that it required students to spend time "playing around" with different types of simulations and code in order to both define and answer their proposed questions. With regard to the collaborative aspect of scientific creativity, direct collaboration on the project was optional; however, the end-of-semester presentations ensured that all students had a chance to share their work with their peers, including time for discussion and questions about each project. The aspect of disciplinary authenticity was addressed in three ways: first, through the explicit use of the computational notebooks, which are a common tool in professional computational physics research; second, by having students engage in the practice of presenting their work to peers in a similar way to professional scientists; and third, in the requirement that students formulate their own research question, rather than working off of a given prompt or question [16].

## III. DATA COLLECTION AND ANALYSIS

In total, 17 students participated in the pilot implementation of the project, working singly or in pairs to produce a total of 11 completed computational essays. Three of the participating students were female and 14 were male; all female students worked in a pair with a male partner. All essays were collected, and all but one pair (15 of 17) students also consented to be interviewed shortly after completing their essays, resulting in a total of about 10 hours of interview data. Interview prompts asked students to walk the interviewer through the development process of their essay, reflect on the connection between computation and learning physics, and reflect on the ways in which computation related to creativity in the courses they were taking. To probe their views of creativity and computation, students were prompted with a statement such as "One of the goals with this project was to give you a little bit of creative freedom, to let you 'play around' with the physics and programming that you have learned so far. Could you talk a little bit about how that felt to you?" However, some students also spontaneously brought up themes related to creativity at various points in the interviews. Since students were native Norwegian-speakers interviewees were given to option to speak English or Norwegian depending on their preference, and three groups chose to conduct the interview partially or entirely in Norwegian.

After all interviews were completed, the resulting recordings were thematically analyzed and coded for themes around students' perceptions of the relationship between computation and physics learning, their general approaches to the computational essay project, and their views on creativity and computation. Essays were also analyzed, with an eye to places where they overlapped with self-reported data from the interviews and places where they addressed things not discussed (like specific coding practices, communication styles, and report structures). This analysis contributed to the emerging themes.

In what follows, we address our second research question by presenting the subset of emergent themes specifically related to computation and creativity.

## IV. RESULTS

Based our analysis, we identified three primary themes: first, that students felt that this type of open-ended programming-based project presented numerous opportunities for creative freedom on both a macro and micro-level; second, that they felt this freedom was a major source of challenge in the project; and third, that they felt this freedom was motivating. Below, we elaborate on these themes with selected quotes from the interview data. All quotes are from English interviews, although these themes were also present in Norwegian interviews, and all names are pseudonyms.

### A. Computation allows for creative freedom in both approach and project

There was a consensus among the interviewed students that programming provides significant creative possibilities, especially in physics. For example, one student, Mel, had this to say:

**Mel:** *So, overall I think programming is a very nice way to be creating, because it's very easy to get results from your creativity. You can do something for, like, not a long time, and then you get something! It's like, it's something specific, and you can change something, and then you can get something else right away. So, it's a, like a very op—a very easy platform to be creative on, generally.*

Here, Mel is explicitly reflecting on the creative affordances of programming, suggesting that it allows you to get interesting results with a relatively small amount of effort. Furthermore, it is extremely easy to modify a written program, and any changes made will have an immediately perceivable effect.

Beyond the general creative affordances of programming, students distinguished between two specific types of creative freedom: freedom on the micro-level, in that programming allows for variation in how individual students accomplish similar sub-tasks, and freedom on the macro-level, in that computation allowed them to pursue a wide variety of topics and projects. On the topic of this micro-level of creativity, another student, Jeffry, added the following:

**Jeffry:** *There were always many different ways to solve a problem. And, in time, thinking about, "Okay, what is the correct syntax in—what is the preferred syntax in doing this? Like is it... Should you write it like that or like that?"*

In other words, programming allows for multiple avenues to accomplish similar tasks, giving students freedom to choose their preferred approach.

In contrast to this micro-level of creativity, another student, Morten addressed the larger-scale creative uses of programming:

**Morten:** *I guess that's the fun part of this, this project as well. Like the computational essay. That you have a kind of a blank slate, you have kind of a situation you want to explore, and then you, like, make your own problem, so to speak. And then kind of just see what happens. And that's a lot easier to do using programming.*

Here, Morten echoes Mel's sentiment that programming is a useful tool for inquiry because it allows you to quickly make modifications and see the effects of those on your simulation. However, he also connects this affordance to the open-endedness of the project, suggesting that it allows you to "make your own problem."

These reflections are supported by the fact that, across all 11 projects, no two groups chose exactly the same topic and approach. There was a roughly even spread of projects based on the various pre-built simulations provided, but in most cases students who chose the same topic diverged significantly in their implementation and investigation questions. For example, of the two groups who investigated cyclotrons, one did a straightforward implementation of the effects of special relativity on the provided simulation, while the other re-wrote the simulation to investigate the effects of relativity on the LHC.

This theme provides confirmation that our attempt to design an open-ended project based on the principles of scientific creativity (research question 1) was at least partially successful.

### B. Students found the openness of the project challenging and appreciated the scaffolds

Although students expressed appreciation for the openness in the project, they also reported that this openness was a significant challenge in completing the project. For example, Jeffry added the following in his interview:

**Jeffry:** *We had the freedom to choose, of course. The problem is that it makes it hard to choose a topic. So I went back and forth, looked at different topics. And even once I'd chosen that I wanted to do, the cyclotron, also trying to think of, 'what is it I want to do with a cyclotron?'*

Gerald echoed this sentiment in his interview:

**Gerald:** *I think, from where we were in the very beginning, [it] was a bit hard because it was almost a bit too open, 'Cause I didn't really know what to do. And so maybe we spent a bit too much time playing around with it and stuff.*

In these two excerpts, both Jeffry and Gerald point to the difficulty of identifying an interesting project topic. This difficulty is unsurprising, since the process of identifying a "good" research question is a challenging prospect, regardless of field [16]. Thus, we argue that this challenge is authentic to the discipline of physics, likely more so than the standard difficulties students face on problem-solving assignments.

Despite these challenges, the students expressed appreciation for the provided guidelines and scaffolds, suggesting that they helped to temper this difficulty. For example, Morton's partner Kurtis added the following:

**Kurtis:** *It was good to have something to work around. As a starter. When, like, it was like, 'okay, use electromagnetics to do something' and then we were like 'okay, what does that mean?' It's very good to have these examples to begin with.*

Here, Kurtis refers to the fact that the original project description left the topic very open, essentially asking students to "investigate a phenomenon related to electricity and magnetism." Although the students found this openness challenging, they appreciated being able to use the provided simulations as a starting point for their investigation.

### C. Students were motivated by this creative freedom

Although they found it challenging, most of the students also reported that they felt this freedom was motivating, favorably comparing the computational essay project to more standard programming-based assignments. For example, Lily made the following comparison:

**Lily:** *Honestly I think this is better than the obligs [obligatory assignments] in a way because I think we pushed ourselves harder here than we would with those assignments. Because then you have an endpoint like, okay, I know what the program or what the assignment asked me to do and here's the program. But now, when we finished something, it was like 'this is really cool to actually see. What else can we do?'*

Here, Lily explicitly compares this project to compulsory weekly assignments, stating that she felt motivated by the fact that there was not a defined goal or end-point with this assignment. Mel echoed this sentiment in his interview:

**Mel:** *I feel like I had a lot of creative freedom, and it was fun to, you know—normally, I'm not that easily motivated when I have, like, one thing, and you have to do this specific thing. But now I could choose my own thing and do that thing. So it's much easier to work and be effective when actually working. And it was interesting trying to search on the internet and find more out about this, things I didn't know, and also try to implement parts of the course into the program.*

Mel explicitly states that he is not usually motivated by standard assignments that ask him to do "this specific thing." However, like Lily, with the computational essay project he felt motivated to both work on his project and try to find new features to build into it, drawing inspiration from course topics and online sources.

These reflection are supported by the fact that about half of the students used significantly more time on the computational essay project than would have been expected for the project it replaced. That is, the mandatory assignment supplanted by the computational essay project was expected to take roughly 4-8 hours of work, with some students anecdotally putting in substantially less. However, when the interviewed students were asked how long they spent on computational essays, all groups reported spending over 6 hours on the project, and half (5 of the 10) reported spending upwards of 14 hours per person. Most of the students also reported doing significant background reading and research to find published results with which to compare their simulations, including reading through published scientific literature.

This motivational aspect, we propose, is one of the reasons it is especially important to build more of these types of opportunities for scientific creativity into the physics curriculum. Several students explicitly stated that they felt unmotivated by standard physics assignments; these types of students, we suggest, are being underserved by the standard model of physics instruction. Although activities like open-ended labs, inquiry-based teaching, and undergraduate research experiences may help, we also see open-ended programming projects like this one as an excellent way to bring in additional creative opportunities.

## V. LIMITATIONS AND CONCLUSIONS

Based on the limited nature of this pilot project, we acknowledge that there are significant limitations to these findings. For starters, our pilot was run with a relatively small number of students, making all conclusions preliminary. Another important limitation is the effect of self-selection: because participation in this project was optional, our participants were likely to be the students most familiar and comfortable with programming, as well as those most predisposed to open-ended projects. We also suspect that there was some gender-based self selection, since only 17% of the students in the pilot (3/17) were female, while the overall distribution in the course was closer to 26%. However, although such self-selection is certainly an important factor it is worth noting that over half of the interviewed students reported that they had encountered programming for the first time at the University of Oslo. Additionally, we feel that self-selection has little impact on our argument that these students are likely to be underserved by a traditional physics curriculum.

In the end, we see great promise in both computation generally and computational essays in particular as ways to motivate students and open up room for creativity in the physics curriculum. Based on this pilot, during the upcoming semesters we plan to expand this project to be a mandatory part of the full electricity and magnetism course and will be collecting larger-scale data on both the variety of projects students produce and the types of guidelines and supports needed to run this kind of project. We are also investigating how computational essays can support students' computational thinking skills and computational literacy in physics [9]. In the years to come, we hope to see computational essays emerge as an important communication tool within the discipline of physics.